\documentstyle[12pt,epsfig]{article}

\newcommand{\zit}{\bibitem}
\newcommand{\proc}[2]{\mbox{$ #1 \rightarrow #2 $}}
\newcommand{\X}{\mbox{\em X}}

\newcommand{\Qsquare}{\mbox{$Q^2 $}}

\newcommand{\unit}[1]{\rm \,#1}
\newcommand{\GeV}{\unit{GeV}}

\newcommand{\CC}{\mbox{$CC $}}
\newcommand{\NC}{\mbox{$NC $}}

\newcommand{\Vcc}{\mbox{$p_{T,h}$}}
\newcommand{\dd}{\mbox{d}}

\newlength{\dinwidth}
\newlength{\dinmargin}
\begin{document}
\sloppy
\begin{titlepage}
\begin{flushleft}
%
%
{\tt  DESY 96-046 \hfill ISSN 0418-9833\\
March 1996}

\end{flushleft}

\noindent
\vspace*{3.5cm}
\begin{center}\begin{Large}
%
%
{\bf Measurement of the $\mathbf Q^{2}$ dependence of the
Charged and Neutral Current  Cross
     Sections in $\mathbf e^{\pm}p$ Scattering at HERA \\}
\vspace*{2.cm}
H1 Collaboration \\
\end{Large}
%
%
\vspace*{4.cm}
{\bf Abstract:}
\begin{quotation}
\noindent
The $Q^{2}$ dependence and the total cross sections for charged and 
neutral current processes are measured in $e^{\pm}p$ 
reactions for  transverse momenta of the outgoing lepton larger than 25\GeV.
Comparable size of cross sections for the neutral current 
process and for  the   weak charged current process are
observed above  $Q^2~\approx~5000$~GeV$^2$. Using
the shape and magnitude of  the charged current  cross section we 
determine a propagator mass  of  $m_{W} = 84\ ^{+10}_{-7}$ GeV. 
\end{quotation}
\vfill
%
%
\end{center}
\end{titlepage}
\footnotesize\noindent
 S.~Aid$^{14}$,                   
 V.~Andreev$^{26}$,               
 B.~Andrieu$^{29}$,               
 R.-D.~Appuhn$^{12}$,             
 M.~Arpagaus$^{37}$,              
 A.~Babaev$^{25}$,                
 J.~B\"ahr$^{36}$,                
 J.~B\'an$^{18}$,                 
 Y.~Ban$^{28}$,                   
 P.~Baranov$^{26}$
% ...

\newpage\footnotesize
\noindent
% ...
$\:^2$ III. Physikalisches Institut der RWTH, Aachen, Germany$^ a$ \\
% ...
$ ^{12}$ DESY, Hamburg, Germany$^a$ \\
% ...
$ ^{14}$ II. Institut f\"ur Experimentalphysik, Universit\"at Hamburg,
                                     Hamburg, Germany$^ a$  \\
% ...
$ ^{18}$ Institute of Experimental Physics, Slovak Academy of
                Sciences, Ko\v{s}ice, Slovak Republic$^ f$\\
% ...
$ ^{25}$ Institute for Theoretical and Experimental Physics,
                                                 Moscow, Russia \\
 $ ^{26}$ Lebedev Physical Institute, Moscow, Russia$^ f$ \\
% ...
$ ^{28}$ LAL, Universit\'{e} de Paris-Sud, IN2P3-CNRS,
                            Orsay, France\\
 $ ^{29}$ LPNHE, Ecole Polytechnique, IN2P3-CNRS,
                             Palaiseau, France \\
% ...
$ ^{36}$ DESY, Institut f\"ur Hochenergiephysik,
                              Zeuthen, Germany$^ a$\\
 $ ^{37}$ Institut f\"ur Teilchenphysik,
          ETH, Z\"urich, Switzerland$^ i$\\
% ...
$ ^a$ Supported by the Bundesministerium f\"ur Bildung, Wissenschaft,
        Forschung und Technologie, FRG,
        under contract numbers 6AC17P, 6AC47P, 6DO57I, 6HH17P, 6HH27I,
        6HD17I, 6HD27I, 6KI17P, 6MP17I, and 6WT87P \\
% ...
$ ^f$ Supported by the Deutsche Forschungsgemeinschaft\\
% ...
$ ^i$ Supported by the Swiss National Science Foundation\\

\newpage \normalsize
%
\section{Introduction}
%
The unification of weak and electromagnetic forces is an important ingredient
 of the Standard Model.
As a consequence  the  cross sections for neutral current (\NC) and
charged current (\CC) processes
\begin{equation}  
\proc{e^{\pm}p}{e^{\pm}+ \X} \mbox{\rm~~~~~and~~~~~} 
{e^{\pm}p} \rightarrow\stackrel{\small (-)}{\nu}_e + \X 
\end{equation}
($\X $ denotes the hadronic system recoiling 
against the lepton)  
 are predicted to be of comparable size when
probing small distances.
  The H1 experiment at the electron--proton  collider HERA permits
  the study of these processes at  large squared four momentum transfer $Q^2$
using the same detector  for the  measurement of both processes.

The generalized cross sections for deep--inelastic $e^{\pm}p$-scattering 
  can be written as \cite{gsf}:
\begin{eqnarray}
\label{eq:wqnc}
\frac{\dd^{2}\sigma(e^{\pm}p)}{\dd x \dd Q^2} = 
\frac{2\pi\alpha^2}{xQ^4}\left((1+(1-y)^2){\cal F}_2 \mp
(1-(1-y)^2)x {\cal F}_3\right).
\end{eqnarray}
Here $x$ and $y$ are the Bjorken scaling variables which are related
by $Q^2=sxy$. The center of mass energy squared $s$ is given by the
product of the electron and proton beam energies $s=4E_eE_p$. The 
electromagnetic coupling constant is denoted $\alpha$ 
 and ${\cal F}_{2,3}$ are generalized
structure functions.  These cross sections contain not 
only the proton structure 
functions but also the electroweak coupling constants and 
propagator terms.  The
 contribution of the longitudinal structure function
$F_{L}$ is expected to be small at high $Q^2$ and is neglected here.

The \NC\  cross section is governed by the 
photon propagator while the $Z^0$ exchange 
affects  the cross sections significantly only at $Q^2 \approx
m_{Z^0}^2$. 
Also the 
contribution of the $Z^0$ exchange enters with different 
signs for $e^+p$ and $e^-p$ scattering. The difference between the two
\NC\ cross sections  therefore provides a method for detecting these 
electroweak effects. However, the accumulated statistics of the
\NC\ data are not yet sufficient to distinguish the dominant photon 
contribution from $Z^{0}-$exchange. 

 For \CC\ processes the lowest order cross section  can be
simplified to
\begin{eqnarray}
\label{eq:wqcc-}
\frac{\dd^{2}\sigma_{\rm CC}^{e^-p}}{\dd x \dd Q^2} = 
 \frac{G_{\mu}^2}{2\pi}\frac{1}{(1+Q^2/m_{W}^2)^2}
\left[ \sum_{q=u,c} q(x,Q^2) + \sum_{\bar q=\bar d,\bar s, \bar b} (1-y)^2 \bar{q}(x,Q^2)\right]
\end{eqnarray}
\begin{eqnarray}
\label{eq:wqcc+}
\frac{\dd^{2}\sigma_{\rm CC}^{e^+p}}{\dd x \dd Q^2} = 
\frac{G_{\mu}^2}{2\pi}\frac{1}{(1+Q^2/m_{W}^2)^2}\left[
 \sum_{\bar q=\bar u,\bar c}\bar{q}(x,Q^2)+\sum_{q=d,s,b}(1-y)^2 q(x,Q^2)\right]
\end{eqnarray}
where $q$ and $\bar{q}$  are the quark  and anti-quark densities of 
the proton. $G_{\mu}$ is the
Fermi constant known from the measurement of the muon lifetime.
In the propagator term $m_{W}$ denotes the mass of the  $W$ bosons
which mediate  the weak force.

In our previous publications total charged current cross section
measurements  have been presented   \cite{ew1,ew2}.
In this analysis we
study the dependence of the  charged current cross section
on  the four momentum transfer. Such an analysis has been presented 
earlier by the ZEUS Collaboration \cite{zeus} for 1993 $e^-p$. Here we present 
results obtained from both  $e^-p$ and  $e^+p$ data recorded in 1993 and 1994.
From the cross section and  the shape of the  
$Q^2$ distribution we derive $m_{W}.$
  
Neutral current cross sections at  high     $Q^2$ have already been 
precisely
measured using the information provided by the scattered lepton only
\cite{ci} or together with the hadronic final state \cite{f2}.
For the comparison with charged current
processes we use here the hadronic final state for event selection
as well as for kinematic reconstruction. This procedure reduces 
uncertainties arising from different systematic effects.

\section{Experimental conditions}
%
 HERA is an  $ep$-collider where  electron or positron beams
  collide with 820 GeV protons.
This analysis is based on data taken in 1993
and 1994 using the H1 detector.
The integrated luminosities are 0.33 pb$^{-1}$ (1993) and 
 0.36 pb$^{-1} $(1994) for $e^-p$   running  and 
2.70 pb$^{-1} $(1994) for  $e^+p$ running.
The lepton beam energies were 26.7~GeV in 1993 and
27.5~GeV\   in 1994.

%
H1 is a multi--purpose detector well optimized to measure large $Q^2$ events in
deep inelastic scattering.
A detailed description of the H1 detector and its performance can be
found in \cite{H1}. We discuss  those aspects which are relevant to the present
analysis.

A system of central and forward
drift chambers is
 used to determine the interaction  vertex of  an event. The chambers measure 
tracks from charged particles in the polar angular range of $7^\circ
-165^\circ$ with respect to the proton beam 
direction (which is the $+z$-direction).
The drift chambers are complemented by
several layers of proportional chambers which  provide
 fast trigger signals.
 
The energy and angles of the  hadronic final state particles as well as
the scattered electron are 
 measured in the highly segmented ($\approx 45000$ cells)
liquid argon (LAr) calorimeter  \cite{LAR78}.  
It covers  a polar angular range 
between $4^\circ$  and $153^\circ$. 
The total depth of the  calorimeter varies between 4.5 and 8 interaction
lengths.
The resolution for single pions and electrons 
is $\sigma(E)/E \approx 0.50/\sqrt{(E/\mbox{GeV})} \oplus
0.02$ and
 $\sigma(E)/E \approx 0.11/\sqrt{(E/\mbox{GeV})} \oplus 0.01$ 
 respectively, as measured in test beam \cite{H1PI}.
The hadron energy scale is  verified within
 $4\%$ with data taken at HERA \cite{ptbal}.
  For trigger purposes the
 LAr calorimeter is read out in a  coarser granularity via
flash ADCs. They provide  a fast  measurement of energies and allow 
triggering on missing  transverse momenta (\CC\ trigger) or on
large electromagnetic energy deposition (\NC\ trigger).
  
Both the LAr calorimeter and the central  drift chambers are surrounded by a
super-conducting solenoid. The iron  yoke which returns the magnetic flux
is instrumented with streamer tubes and is used as a muon detector.

A luminosity system measures the rate of the small angle  Brems\-strahlung
process $ep~\rightarrow~ep\gamma$ \cite{lumi} and allows the
determination of   the luminosity  with a precision of
1.5\% (3.5\% in 1993).

\section{Data Samples}
%
The selection starts from all triggered  events and 
is based on information of the hadronic final state both for
  \CC\ and \NC\ interactions.
  The presence or absence of a charged scattered lepton 
is used only to classify  an event as a candidate for a \CC\ or a \NC\ 
process. 

Scattered electrons are identified in the LAr calorimeter by the
lateral extension and energy density  of the electromagnetic shower and
the requirement that a reconstructed track points to the
shower.  The minimal transverse energy  deposition of the electron 
candidate
is required to be  5 GeV \cite{andre}.
The search is restricted to scattering angles
of the charged lepton $\vartheta_{e} < 150^o$, which is well inside the
angular acceptance of the liquid argon calorimeter.
The   efficiency of the electron identification is   determined by 
comparing   independent 
electron   identification algorithms, either  based on calorimeter information 
only or on a track criterion.   
 A total electron identification 
 efficiency  of $98.7 \pm 0.3 \%$ is found for the finally analysed \NC\ sample.

After the  neutral current candidates are identified
 all information about the electron in the
track detectors and calorimeters is ignored in the subsequent  analysis.  
The additional selection described below is applied to both samples on 
the hadronic final state only.
 The selection steps are as follows:

\begin{enumerate}
\item{The  hadronic  transverse momentum sum $p_{T,h}$
 measures the   missing transverse momentum with respect to the incoming particle 
direction in each event and is
used as a main selection cut requiring
\begin{eqnarray}
\label{VS} 
p_{T,h} = \sqrt{ (\sum_{i}p_{{x}_i})^2 + (\sum_{i}p_{{y}_i})^2} 
> 25 \mbox{\rm ~~GeV} \; .
\end{eqnarray}
 The computation uses  the event vertex position and
energies and positions of LAr  calorimeter cells (index $i$) of the hadronic final state. 
Neglecting particle masses the three components of the momentum vector are then
defined by ($p_{x}, p_{y}, p_{z})$.
}
\item{ For both event types the \CC\ trigger condition has to be 
fulfilled.
For \CC\ events this signal is derived  from the LAr
   calorimeter trigger energies, for \NC\ events it is evaluated  from the
          same measurement after electron subtraction. To assign the events 
          to a specific  crossing of electron and proton bunches,
           a trigger signal from the 
          proportional chambers is required.
}
\item{A vertex is reconstructed from measured tracks.
 This vertex
position has to be within a distance of 35 cm from the nominal
interaction  point along the beam axis. 
}
\item  High energy cosmic or proton beam related muons may cause large
energy deposits in the calorimeter and thus fulfill the  
selection criteria. Based on the topology of these deposits 
 and event timing with respect to the beam 
crossing background filters have been developed which remove these 
events efficiently \cite{ew2}.
\end{enumerate}
The   efficiency for  these four  selection criteria is
determined from the initial \NC\ sample. 
The selection efficiencies are found to be between 80\% and 85\% for the four
processes with an uncertainty of 3\%. All selection efficiencies 
depend only little on kinematic properties of the events.
 The final event
statistics is given in Tab.~\ref{sigmas}.

The \CC\--candidate sample is   subject to a visual inspection  by which
 residual background from incoming  muons 
 and from misclassified \NC\ events
  is identified and removed \cite{ew2}.
  Photoproduction events $\gamma p \rightarrow 
  hadrons$ in which the 
  scattered electron remains undetected  ($Q^{2} \approx 0$) 
  are a potential background for the \CC\ sample if the hadronic final 
  state  is badly
reconstructed due to particle losses in the beam pipe or resolution effects.
 This background has 
  been   estimated from data and was found to be less than $0.5\%$ as described in 
\cite{ew2,andre}. Therefore it has been neglected in  further analysis here.

The total background rate for \NC\ events is
 $2.0 \pm 0.6 \%$ and has been statistically subtracted from the sample.
  The largest background contribution is due to $\gamma p$
events with many hadrons in the final state, one of which is 
misidentified as an electron. For some of these events the scattered 
electron is detected in the luminosity system of the H1 detector. 
From a study of these  tagged events we find a  background 
fraction of $1.4 \pm 0.6 \%$ \cite{andre}. 
Events of the type 
$e^{\pm}p \rightarrow e^{\pm} e^+ e^- X$  have been
found by a visual inspection of the sample. They
contribute 
an additional
 $0.3 \pm 0.2 \%$ to the background.
  The background from incoming muons amounts to
$0.3 \pm 0.1 \%.$

%
\section{Data Analysis}
The final state kinematics is fully specified by  two independent 
variables. 
Following the method proposed in \cite{JB} we use
the   transverse momentum sum \Vcc\ (cf.\ eq.\ (\ref{VS})) 
and the energy transfer  $y_{h}$:
\begin{equation}
y_{h}=\sum_i (E_i - p_{{z}_{i}})/2E_e.
\end{equation}
Here also particle masses are   neglected and the summation is performed 
 over all LAr calorimeter cells $i$ of the hadronic final state.

 For comparison to data and for studies of migration effects and unfolding
  we use a detailed detector simulation. The electroweak processes are 
  generated by the DJANGO Monte Carlo program that includes  QED radiative effects of
  order ${\cal O} (\alpha)$
\cite{Django}  with proton parton densities of the MRSH 
parameterization \cite{MRSH}. The program makes use of the color dipole 
model as implemented in  ARIADNE \cite{aria} for hadronization.

In Fig.\ 1 the measurements of \Vcc\ and $y_{h}$  are compared with
 the expectations from the simulation  for
the \NC\ sample.
The simulation agrees well with  the measured distributions   within the
statistical errors. 

\begin{figure}[htb]
   \centering 
   \epsfig{file=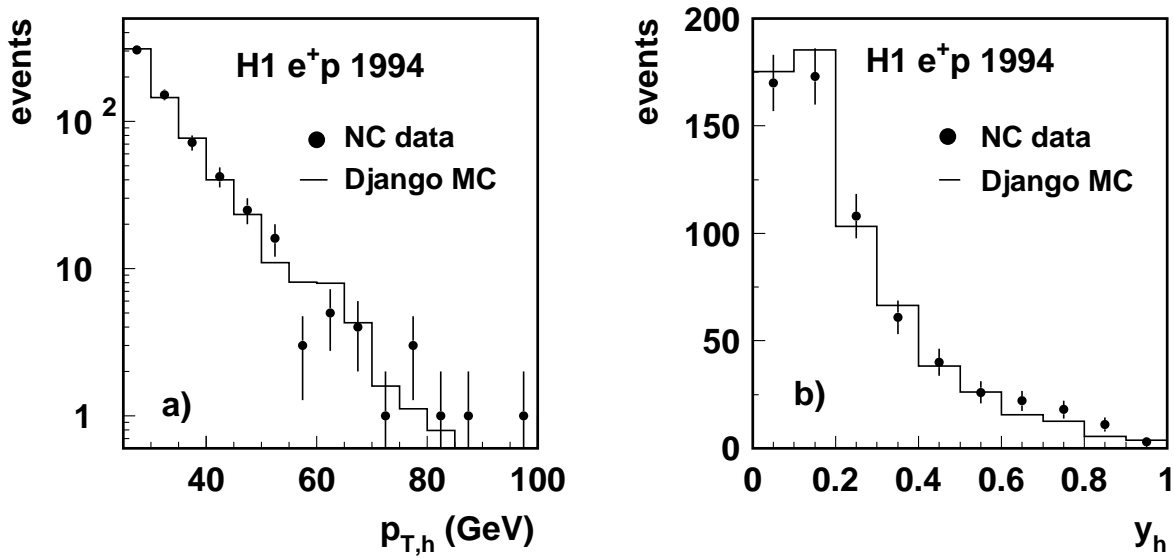,width=0.99\textwidth}
   \caption{\small \sl
Neutral current $e^{+}p$ data (filled circles) are compared to
      a Monte Carlo (MC) simulation (histogram) normalized to the observed number of events.
 Shown is the transverse
      momentum sum $p_{T,h}$  of the hadronic final state is shown in
      (a) and  $y_{h}$ in (b) . 
   For the data only statistical errors are given. The statistical errors from
 the MC simulation are negligible.}
   \label{yv} 
\end{figure}
The squared four momentum transfer
\Qsquare\ is related to $p_{T,h}^2$ and $y_{h}$ by
\begin{equation}
 Q^2_{h} = {  {p_{T,h}^2} \over{1-y_{h}}}.
\end{equation}
The relative resolution for $Q^2_{h}$ is about 20\%.
For experimental cross checks
the  measurement of  $Q^2_e$ by the scattered electron in \NC\
events can be compared with $Q^2_{h}$ determined from the hadronic
 final state.
The lepton scattering angle $\vartheta_e$ and energy
$E'_e$ are used for calculating  the kinematics with:
$Q^2_e = 2 E_e E'_{e} (1+\cos (\vartheta_{e}))$. 
 The  $Q^2_e$ measurement has a 
resolution  $\Delta Q^2_{e}/Q^2_{e} \approx3\%$ . In
Fig.~2  we compare  the measured ratio  $Q^2_{h}/Q^2_e$ with the
 Monte Carlo simulation. The comparison
shows that the hadronic measurement of the kinematics is well 
described by the simulation.
The distribution is not centered around one due to losses in transverse 
momentum of the hadronic final state into the beam pipe.

\begin{figure}[htb]
   \centering 
   \epsfig{file=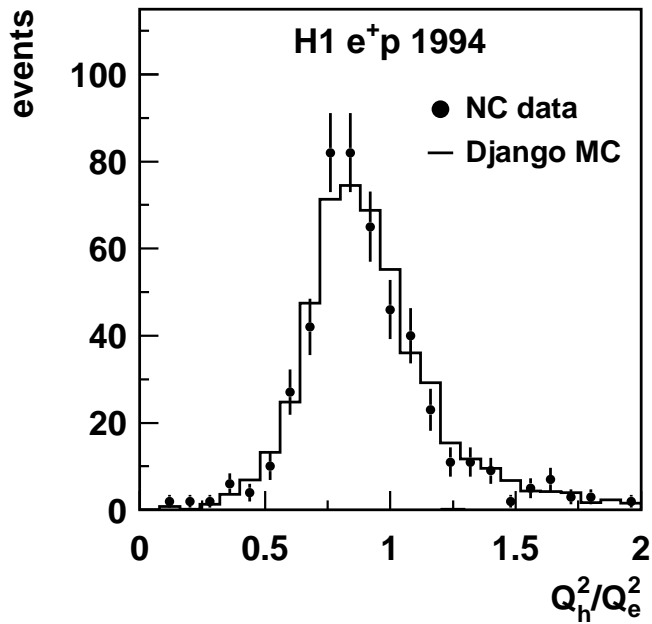,width=10cm}
   \caption{\small \sl The ratio of the squared four momentum 
   transfer as determined from the 
   hadronic final state $Q^2_{h}$ and from the  scattered electron  
   $Q^2_e$ for \NC\ $e^{+}p$ events compared to the Monte Carlo prediction.
   For the data only statistical errors are given. The statistical errors from
 the MC simulation are negligible. }
   \label{q2hade} 
\end{figure}

We apply a matrix unfolding technique \cite{zech} to determine the 
true $Q^2$ distribution from the measured $Q^2_{m}$ distribution.
 The true squared four momentum $Q^2$ is  defined by the  
outgoing lepton. For the unfolding we make 
use of the relation
\begin{equation}
  \frac{ d\sigma(Q^{2}_{m})}{dQ^2_{m}} = \int dQ^2 A(Q_{m}^2,Q^2)
  { { d\sigma({Q^2})}\over{dQ^2 }}.
\label{fold}
\end{equation}
 $A(Q_{m}^2,Q^2)$ is the detector response function averaged over the 
 variable $x$. Consequently it has to be determined separately for  
 the different 
 processes under study. 
For finite  $Q^2$ bins  equation (\ref{fold})
can be converted into a sum:
\begin{equation}
\Delta\sigma(Q_{m}^2)_{\mu} = \sum_{\nu} T_{\mu \nu} \Delta\sigma (Q^2)_{\nu}
\label{ufom}
\end{equation}
with bin indices $\mu$ and $\nu$ and the transfer matrix  $T_{\mu \nu}$.
The transfer matrix elements are   determined by  Monte Carlo 
simulation. 
They contain  the transformation of the cut in $p_{T,h}$ into the  
 corresponding experiment independent cut in the true $p_{T}$ of the scattered lepton.
 The calculated transfer matrix is cross checked for \NC\ data 
using  the complementary information provided by the electron.
The binning was chosen such (cf. Tab.~\ref{corrections}),
 that the majority (typically 60\%)
  of the events are reconstructed into the
original  $Q^2$   bin.
Inverting relation (\ref{ufom}) and
applying it to the measured binned distribution gives  the
  true distribution together with its covariance matrix.

In binned distributions the information on individual events is lost.
To exploit the full
information on $m_{W}$ contained  in  the shape of the \CC\
$Q^2$ distribution and the total cross section we use the method of extended
maximum likelihood  \cite{Barlow}.  
The   likelihood function
is defined as:
 \begin{equation}
L = \prod_{i=1}^{n} \frac{P_{i}(Q^2_{m})}{\cal N} \cdot
\frac{e^{-\cal N} \cdot {\cal N}^{n}}{n!}
\label{EMLL}
\end{equation}
where $i$ runs over $n$ individual events.
For an event $i$ the
distribution   $P_i(Q^2_{m})$ in four momentum $Q^2_{m}$ is given 
by the convolution of the
differential cross section $d\sigma(Q^2,m_{W})/dQ^2$ and the detector 
response function  $A(Q_{m}^2,Q^2)$ (cf.\ eq.\ (\ref{fold})). 
The predicted  number of events for a propagator mass $m_{W}$ and
an integrated luminosity $\cal L$ is  ${\cal N}= {\cal L} \int_{}^{} P (Q^2_{m}) dQ^2_{m}.$ 
The propagator mass is then extracted from the data by maximizing the 
likelihood function.

%
\section{Results} 
In Fig.~3 and  Tab.~\ref{sigmas} we present the result of the unfolding of the 
  $Q^2$-differential cross sections
for the $e^{\pm}p$ data  for   \NC\ and \CC\ for 
$p_{T} > 25$ GeV. The theoretical prediction  
was obtained with
the HERACLES  \cite{heracles}  generator using the  MRSA${'}$ \cite{MRSH} 
proton 
parton densities. The uncertainty of the prediction is about 4\% due to
the imperfect knowledge  of the parton densities.
Note that the data are not corrected for radiative effects. 
Born cross sections can be  obtained using                       
$d\sigma_{Born}/dQ^2=d\sigma/dQ^2(1 +  \delta_{Born})$ 
(cf.~Tab.~\ref{corrections}).
A factor $c_{p_{T}}$ which can be used additionally to correct the measured cross 
section to the full phase space $d\sigma/dQ^{2}(p_{T}>0$~GeV$^2)=
c_{p_{T}}  d\sigma/dQ^{2}(p_{T}>25$~GeV$^2)$ is given in
 Tab.~\ref{corrections}. The error on $c_{p_{T}}$ accounts for uncertainties in the
structure functions.
 Results are given at  fixed $Q^2$ values, therefore the measured bin 
 averaged cross section has been corrected by the factor $c_{ bin} = 
 \frac{d\sigma^{theor.} (Q^{2})}{dQ^{2}}/\frac{\Delta \sigma^{theor.}}{\Delta Q^{2}} $ 
 which is given in Tab.~\ref{corrections}. 
 Differences in center of mass  energy  for different running periods
are  taken into account and the results are given for $s=90200$ GeV$^{2}.$

 Agreement
 with the Standard Model prediction is observed.
Our result for \NC\ is also in  agreement with the differential cross section
measurements based on essentially the same data using
 the electron variables  presented in \cite{ci} with slightly 
 different selection criteria.
Above $Q^2~=~5000$~GeV$^2$ the cross sections for neutral and charged current
processes are of comparable size.
  For the $e^+p$ data the charged current cross section stays
 below the neutral current cross section over the whole $Q^2$ range, due
to the different $y-$dependence of the 
quark contributions to the cross section (cf.\ eq.\ (3,4),
\cite{ew2}).

The \CC\ cross section errors  are dominated  by the
statistical errors whereas for the \NC\ samples the main contribution 
to  the errors comes from the uncertainty in the hadronic energy scale.
Other sources of systematic errors are uncertainties in the 
determination of the trigger and selection efficiencies as well as the
luminosity measurement.

  For the neutral current cross section we  show in Fig.~3 the Standard Model
   expectation for $\gamma,Z^0$ exchange and assuming 
photon exchange only. This illustrates
that the present integrated luminosity does not yet allow the distinction 
between the pure photon exchange and the total electroweak cross 
section in \NC\ reactions.

\begin{figure}[p]\centering
\epsfig {file=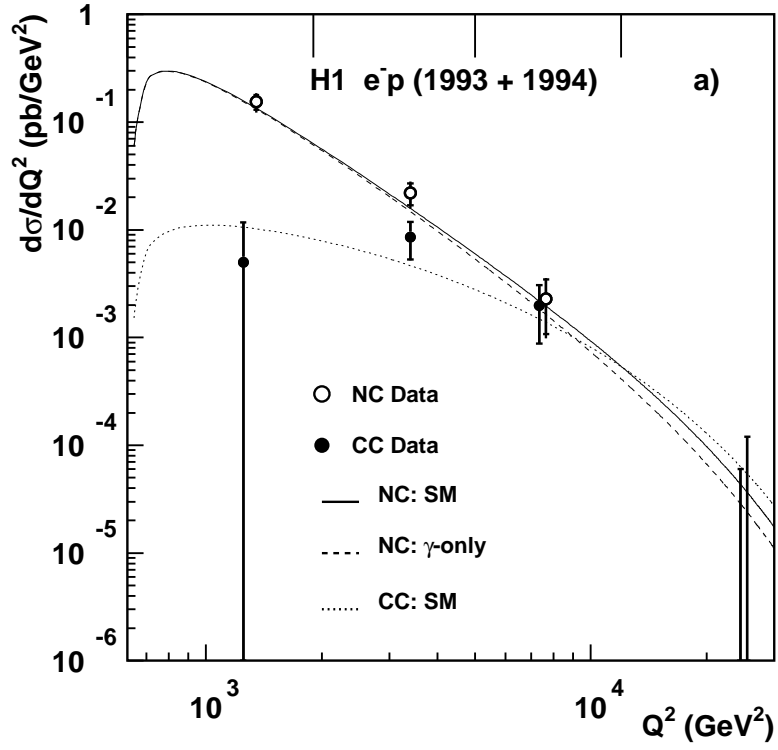,height=.43\textheight}
\epsfig {file=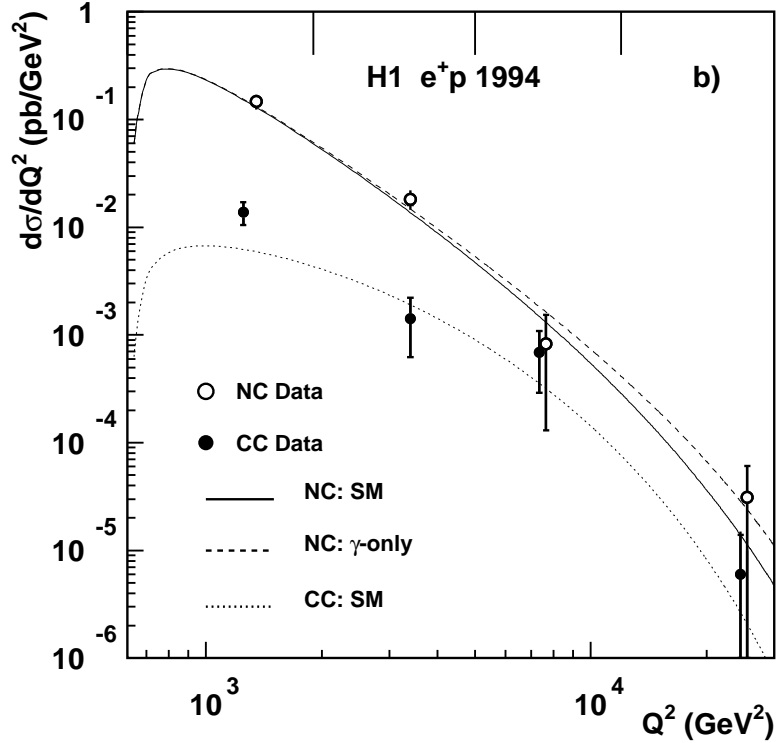,height=.43\textheight}
\caption{\small \sl
        The unfolded differential cross sections $d\sigma/dQ^2$  
         are shown for charged and neutral
        current interactions with $p_{T} > 25$ GeV 
        for $e^-p$ collisions in a) and for $e^+p$ collisions in b).
        The inner error bars contain statistical, the full error bars 
         include also the systematic errors added in quadrature.
        Boundaries of each bin are denoted by the vertical  bars on
        the top of the plots. 
        The solid lines indicate the Standard Model predictions for 
        neutral current interactions and the dotted lines  for  charged current
interactions. Dashed
        lines show the prediction for \NC\ from photon exchange alone.}
\label{mfig}
 \end{figure}

\newcommand{\des}{\pm\delta_{\rm syst.}}
\newcommand{\dec}{\pm\delta_{\rm syst.}}
\newcommand{\dst}{\pm\delta_{\rm stat.}}
\newcommand{\sq}{d\sigma/dQ^2 \over fb/GeV^2}
\newcommand{\db}{\delta_{Born}}
\footnotesize

\begin{table}[htb]\centering
\renewcommand{\arraystretch}{1.6}
  \begin{tabular}{|l|c|c|c|c|c|} 
  \hline
 \multicolumn{2}{|c|}{ } & \multicolumn{2}{c|}{  $e^-p$} & \multicolumn{2}{c|}{  $e^+p$} \\
   \hline
    
$Q^{2} \over GeV^{2}$& &  \NC\           &  \CC\          &  \NC\
&\CC\             \\
\hline
\hline
1250 & events   & 105               & 6                   &  370           & 26  \\ \cline{2-6}
                   & $\sq$    & $155\pm24\pm21$   & $5.0\pm6.7\pm0.9$   &$146\pm12\pm20$ & $13.7\pm3.3\pm0.7$\\  \hline
\hline
3400 & events   &  42          &14            & 133      & 13            \\ \cline{2-6}
                   & $\sq $   &$21.9\pm5.0\pm3.6$ &$8.6\pm3.3\pm0.7$ &$18.2\pm2.2\pm3.0$& $1.4\pm0.8\pm0.2$ \\   \hline
\hline
7500 & events    & 6            & 5           & 16        & 7             \\ \cline{2-6}  
                  & $\sq  $   & $ 2.3\pm1.2\pm0.4$&$2.0\pm1.1\pm0.3$ &$0.8\pm0.7\pm0.2$& $0.7\pm0.4\pm0.1  $ \\   \hline
\hline
25000& events    & 0            & 0           & 2          & 1             \\ \cline{2-6}  
                  & $\sq $    & $0.00\pm0.12\pm0.01$ &  $0.00\pm0.06\pm0.04$&$0.03\pm0.03\pm0.01$&$0.006\pm0.008\pm0.003$\\   \hline 
\hline
  total&   events   & 153            &         25   &            521 &   47       \\ \cline{2-6}
      &$\sigma /  pb$&$331\pm27\pm47$&$46\pm9\pm3$&$299\pm13\pm42$&$23\pm3\pm2  $ \\ \cline{2-6} 
      &$\sigma_{SM} / pb  $ &   $ 294\pm12 $  &  $ 40.7\pm1.6 $   &  $  
      276\pm11 $   &  $ 15.6\pm0.6 $ \\ \hline

  \end{tabular}
  \caption{\small \sl Event statistics, differential, and total cross sections for 
   $p_{T}>25$ GeV are listed for  \NC\ and \CC\
    processes in $e^+p$ and $e^-p$ running. The first error is of 
    statistical nature and obtained from the diagonal element of the
    covariance matrix, the second accounts for systematic uncertainties.
    The total cross sections are compared with the Standard Model 
    prediction for $m_{W}=80.2$ GeV and using the MRSA${'}$ parameterization 
    for the proton parton densities.
 }
  \label{sigmas}

\end{table}

\normalsize

From the \CC\ sample we obtain using the extended maximum likelihood fit 
 a propagator mass of 
 $$
 m_{W}(e^-p) = 78\ ^{+11}_{-9}\ ^{+4}_{-3} \mbox{\rm ~GeV~,~~~~~~~~~~} 
 m_{W}(e^+p)
 =97\ ^{+18}_{-15}\ ^{+5}_{-10}\mbox{\rm ~GeV,} 
 $$ 
 where the first errors are of
 statistical and the second of systematic origin. 
 The statistical uncertainties reflect the different sensitivities of 
 the $e^{+}p$ and $e^{-}p$ cross sections to the propagator mass.
 The main source of 
 systematic errors is the hadronic energy scale.   
 The masses for both charge states are
 compatible and a combined fit results in 
$$
m_{W} = 84\ ^{+9}_{-6}\ ^{+5}_{-4} \mbox{\rm ~GeV}.
$$
     Exploiting the $Q^2$ shape alone                                                                             
       gives a consistent but less precise value for the propagator mass.

The shape and cross section analysis exhibits that both
 charged current processes
are consistent with the  exchange of a $W$ boson which has equal mass for
both charge states. This confirms previous results from HERA on
the $W$-mass \cite{ew1,ew2,zeus} with improved precision.
In addition the mass of the intermediate $W$ boson is in  agreement
with the mass determined from direct production of $W$ bosons at 
$p\bar{p}$-colliders \cite{Wmass}.

 In this analysis we use a larger angular coverage for the tracks 
entering the vertex reconstruction compared to our previous analysis where
 the polar angular acceptance  started  at  $15^\circ$ \cite{ew2}. 
On the basis of this extended acceptance we  update the total \CC\ 
cross sections published in \cite{ew2}.
These cross sections as well as  the ones for \NC\ are collected 
in Tab.~\ref{sigmas}. 
The Standard Model expectations are calculated as before using the HERACLES 
generator, where the error on the expectation originates in the choice of 
the proton structure function.

We present also the ratios of total \NC\ and \CC\ cross sections 
$R=\sigma_{NC}/\sigma_{CC}.$ Due to  
the same treatment of neutral and charged current events many 
systematic errors cancel in this ratio. 
For $p_{T} > 25$ GeV we obtain 
\begin{equation}
R^{+}(e^{+}p) = 12.8 \pm 2.0 \pm 1.0 \mbox{~~~~~~~~~~}R^{-}(e^{-}p) = 
7.2 \pm 1.6 \pm 0.7.
\end{equation}
The result for $R^{-}$ is in good agreement with the theoretical 
prediction of $R^{-}_{SM}=7.2\pm0.1.$ We observe a two 
standard deviation discrepancy for $R^{+}$ from the prediction
 $R^{+}_{SM}=17.7\pm0.9.$
   
 The residual systematic uncertainty is still caused by the hadronic 
 energy scale due to the different cross section slope of \NC\ and 
 \CC\ near the selection cut of $p_{T,h}=25$ GeV. Selecting only 
 events above  $ Q^{2}=5000$ GeV$^{2},$ where the cross section 
 slopes are nearly identical reduces the systematic uncertainty 
 further. For such a cut we obtain $
 R^{+} = 2.0 \pm 1.4 \pm 0.07$ and $R^{-} = 
1.4 \pm 0.8 \pm 0.05,$ while the expected values are
 $R^{+} = 4.17 $ and $R^{-} = 1.34.$
With higher integrated luminosity these 
cross section ratios will be
 sensitive probes for precision tests of the Standard Model \cite{vb}.

\footnotesize
\begin{table}[htb]\centering
\renewcommand{\arraystretch}{1.6}
  \begin{tabular}{|l|l|c|c|c|c|c|} 
  \hline
 \multicolumn{3}{|c|}{ } & \multicolumn{2}{c|}{  $e^-p$} & \multicolumn{2}{c|}{  $e^+p$} \\
   \hline
    
$Q^{2} \over GeV^{2}$&$Q^{2}$ range&   & \NC\           &  \CC\          &  \NC\        & \CC\            \\
\hline
\hline
1250&625--1500   & $c_{bin}$ & 0.956        & 1.191        & 0.950       & 1.213         \\ \cline{3-7} 
          &       & $\db$          & $-0.02\pm0.01$       & 
          $0.06\pm0.02 $       &  $-0.02\pm0.01$    & $0.06\pm0.02 $       \\ \cline{3-7}
 & &$c_{p_{T}}$ &$1.30\pm0.01 $&$1.29\pm0.01 $& $1.30\pm0.01$&$ 1.34\pm0.01$  \\  \hline
\hline
3400&1500--5000  & $c_{bin}$ & 0.703        & 0.990       & 0.674      & 0.939         \\ \cline{3-7} 
          &       & $\db$          & $-0.08\pm0.01$       & $0.04\pm0.02 $     &
           $-0.09\pm0.01$     &$ 0.04\pm0.02 $       \\ \cline{3-7}
 & &$c_{p_{T}}$ &$1.10\pm0.01 $ &$1.10\pm0.01 $ &$1.10\pm0.01 $ &$1.11\pm0.01 $  \\  \hline
\hline
7500&5000--12000 & $c_{bin}$ &  0.949       & 1.129       & 0.894      & 1.099         \\ \cline{3-7}  
          &       & $\db$          & $-0.09\pm0.01$       & $ 
          0.02\pm0.02$    & $-0.12\pm0.01$     & $0.02\pm0.02 $       \\ \cline{3-7}
 & &$c_{p_{T}}$ &$1.06\pm0.01 $ &$1.06\pm0.01 $ &$1.05\pm0.01 $ 
 &$1.07\pm0.01 $  \\  \hline
\hline
25000&12000--s     & $c_{bin}$ & 1.142         &  1.47           & 0.729     & 0.667         \\ \cline{3-7}  
         &         & $\db$          & $-0.07\pm0.01$      &  
         $-0.02\pm0.02$           &  $-0.19\pm0.01$    & 
         $-0.02\pm0.02  $      \\ \cline{3-7}
 & &$c_{p_{T}}$ &$1.04\pm0.01 $ &$1.04\pm0.01 $ &$1.03\pm0.01 $ &$1.04\pm0.01 $     
  \\ \hline

  \end{tabular}
  \caption{\small \sl For each $Q^{2}$ bin the
      bin boundaries, the applied correction factor for the 
    finite bin size $c_{bin}$ and the correction factors to obtain a
    Born cross section $\db$ and to correct for the full phase space
    $c_{p_{T}}$  are given.
 }
  \label{corrections}

\end{table}

 \normalsize
\section{Conclusion}
 
 The $Q^2$ dependence of the neutral and charged current processes
       have been measured for $e^-$ and $e^+$ beams colliding with protons
       for transverse momenta of the scattered lepton above 25 GeV.
The analysis is based on the properties of the hadronic final  state in
both cases.
The differential cross sections, as well as the total cross sections and their
ratios are in agreement with the Standard Model expectations.
At large  momentum transfers  $Q^2~>~5000$~GeV$^2$
we observe that the  weak and electroweak forces have the same
strength. 
The  propagator masses measured in charged current scattering are
  consistent with being the same for both charges.
  The average is  $m_{W} = 84\ ^{+10}_{-7}$ GeV and is consistent 
  with the on-shell measurement of the $W$ boson mass.

\section*{Acknowledgment}
%
We are grateful to the HERA machine group whose outstanding efforts
made this experiment possible. The immense effort of the
engineers and technicians who constructed and maintained the detector
is appreciated.
We thank the funding agencies for financial support. We acknowledge the
support of the DESY technical staff and wish to thank the DESY
directorate for the hospitality extended to the non-DESY members of the
collaboration.
We thank H. Spiesberger for discussions concerning  radiative corrections
and help with the theoretical calculations.




\end{document}